\pdfoutput=1

\documentclass[11pt]{article}

\usepackage[final]{acl}

\usepackage{times}
\usepackage{latexsym}
\usepackage{enumitem}
\usepackage{multirow}
\usepackage{threeparttable}
\usepackage{hyperref}
\usepackage{CJK}
\usepackage[T1]{fontenc}

\usepackage[utf8]{inputenc}

\usepackage{microtype}

\usepackage{inconsolata}

\usepackage{graphicx}
\usepackage{booktabs}
\usepackage{tabularx}
\usepackage{url}
\usepackage{colortbl}
\usepackage{pifont}
\usepackage{circledsteps}
\usepackage{amssymb}
\usepackage{utfsym}
\newcommand{\error}{\scalebox{0.75}{\usym{2613}}}

\title{CompileAgent: Automated Real-World Repo-Level Compilation with Tool-Integrated LLM-based Agent System}

\author{
 \textbf{Li Hu\textsuperscript{1}},
 \textbf{Guoqiang Chen\textsuperscript{2}},
 \textbf{Xiuwei Shang\textsuperscript{1}},
 \textbf{Shaoyin Cheng\textsuperscript{1,3,*}},
 \textbf{Benlong Wu\textsuperscript{1}},
 \textbf{Gangyang Li\textsuperscript{1}},
\\
 \textbf{Xu Zhu\textsuperscript{1}},
 \textbf{Weiming Zhang\textsuperscript{1,3}},
 \textbf{Nenghai Yu\textsuperscript{1,3}}
\\
 \textsuperscript{1}University of Science and Technology of China
 \textsuperscript{2}QI-ANXIN Technology Research Institute \\ 
 \textsuperscript{3}Anhui Province Key Laboratory of Digital Security
\\
 \texttt{\{pdxbshx,shangxw,dizzylong,ligangyang,zhuxu24\}@mail.ustc.edu.cn} \\
 \texttt{\{sycheng,zhangwm,ynh\}@ustc.edu.cn}  \
 \texttt{guoqiangchen@qianxin.com}
}

\begin{document}
\maketitle

\begin{abstract}

With open-source projects growing in size and complexity, manual compilation becomes tedious and error-prone, highlighting the need for automation to improve efficiency and accuracy. However, the complexity of compilation instruction search and error resolution makes automatic compilation challenging. Inspired by the success of LLM-based agents in various fields, we propose CompileAgent, the first LLM-based agent framework dedicated to repo-level compilation. CompileAgent integrates five tools and a flow-based agent strategy, enabling interaction with software artifacts for compilation instruction search and error resolution. To measure the effectiveness of our method, we design a public repo-level benchmark CompileAgentBench, and we also design two baselines for comparison by combining two compilation-friendly schemes. The performance on this benchmark shows that our method significantly improves the compilation success rate, ranging from 10\% to 71\%. Meanwhile, we evaluate the performance of CompileAgent under different agent strategies and verify the effectiveness of the flow-based strategy. Additionally, we emphasize the scalability of CompileAgent, further expanding its application prospects.


\end{abstract}

\section{Introduction}
\label{sec:Introduction}

Compilation is the process of converting source code into executable files or libraries. Currently, many open-source tool libraries and application software projects can be used directly after compiling into executable files or libraries. Not only that, these files or libraries can also be used for subsequent work, including building diverse datasets \cite{CP-BCS}, conducting performance testing and optimization \cite{OpenCGRA}, security and vulnerability analysis \cite{BinaryAI}, etc.

For single-file compilation, the compiler only needs to process a single source code file and generate the corresponding target code. However, compiling an open-source code repository shared by others is a far more complex, time-consuming \cite{CLAP} and demanding task in actual software engineering. This process goes beyond handling the source code itself and requires addressing intricate challenges such as environment adaptation, dependency management, and build configuration. As a result, developers tend to spend most of their time troubleshooting challenges during the compilation process. 

To date, no research has specifically focused on how to achieve automated compilation at the repository level. Drawing from developers' experience in compiling code repositories, we identify two core challenges in this task. The first is the discovery and accurate extraction of compilation instructions from repositories, which often involve varied build systems, scripts, and configurations. The second challenge is resolving compilation errors encountered during the process, which is required to address issues such as dependency conflicts, environment mismatches, and code compatibility.

Recently, the application of LLM-based agents for automating complex tasks has gained significant attention across various fields. They have been successfully employed in areas such as code generation \cite{agentcoder,CodeAgent}, bug fixing \cite{marscode,repairagent}, and penetration testing \cite{pentestgpt,pentestagent,pentest-ai}, where they autonomously perform tasks that traditionally require human intervention. Inspired by the success of these applications, we propose leveraging agents for the automation of repository-level compilation tasks. By doing so, we aim to streamline the compilation process, reduce manual intervention, and address the challenges inherent in compiling open-source repositories.

In this paper, we propose CompileAgent, the first novel approach that leverages LLM-based agents for automated repo-level compilation. To address the two key challenges identified earlier, we have designed five specialized tools and a flow-based agent strategy. CompileAgent can effectively complete the compilation of code repositories by interacting with external tools. To evaluate the effectiveness of our approach, we manually constructed CompileAgentBench, a benchmark designed for repository compilation. This benchmark consists of 100 repositories in C and C++, sourced from Github. We further conducted comprehensive experiments to evaluate the performance of CompileAgent by applying it to seven well-known LLMs, with parameter sizes ranging from 32B to 236B, to demonstrate its broad applicability. When compared to the existing baselines, CompileAgent achieved a notable increase in compilation success rates across all LLMs, with improvements reaching up to 71\%. Additionally, the total compilation time can be reduced by up to 121.9 hours, while maintaining a low cost of only \$0.22 per project. We compared the flow-based strategy with several other strategies suitable for the compilation task, further validating its effectiveness. Moreover, we conducted ablation experiments to validate the necessity of each component within the system. These experiments provide strong evidence that CompileAgent effectively addresses the challenges of code repository compilation.

Our contributions can be summarized as follows:
\begin{itemize}[left=0.2cm] 
    \setlength{\itemsep}{0pt}  
    \vspace{-0.5ex}
    \item We make the first attempt to explore repo-level compilation by LLM-based agent, offering valuable insights into the practical application of agents in real-world scenarios.
    \vspace{-0.5ex}
    \item We propose CompileAgent, a LLM-based agent framework tailored for the repo-level compilation task. By incorporating five specialized tools and a flow-based agent strategy, the framework enables LLMs to autonomously and effectively complete the compilation of repositories.
    \item We construct CompileAgentBench, a benchmark for compiling code repositories that includes high-quality repositories with compilation instructions of varying difficulty and covering a wide range of topics.
    \item Experimental results on seven LLMs demonstrate the effectiveness of CompileAgent in compiling code repositories, highlighting the potential of agent-based approaches for tackling complex software engineering challenges.
\end{itemize}

\begin{figure*}[t]
  \centering
  \includegraphics[width=0.94\linewidth]{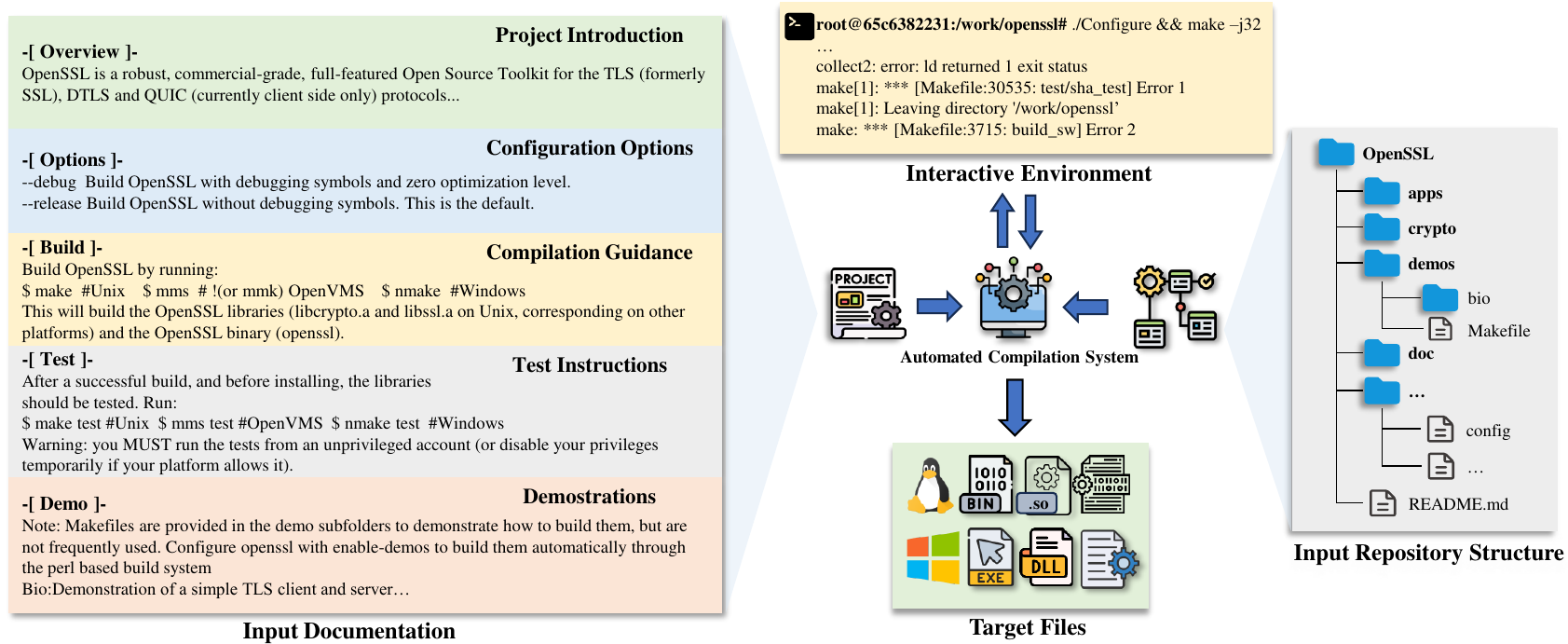}
  \vspace{-1.0ex}
  \caption{An illustrative example of the automated repo-level compilation. The task input contains code repository documentation and structure, and the automated compilation system can interact with the interactive environment.}
  \vspace{-2.5ex}
  \label{fig:automated compilation system}
\end{figure*}

\vspace{-0.5ex}
\section{Background} \label{sec:Background}
\vspace{-0.5ex}

\subsection{LLMs and Agents}
\vspace{-0.3ex}

LLMs have demonstrated remarkable performance across a wide range of Natural Language Processing (NLP) tasks, such as text generation, summarization, translation, and question-answering. Their ability to understand and generate human-like text makes them a powerful tool for various applications. However, LLMs are limited to NLP tasks and struggle with tasks that involve direct interaction with the external environment.

Recent advancements in LLMs have significantly expanded their capabilities, with many models now supporting function calls as part of their core functionalities. This enhancement allows LLMs to dynamically interact with external systems and tools, playing a key role in the development of the AI agents \cite{tellmemore,mapcoder,queryagent,chatdev,chen2023towards,xie2023openagentsopenplatformlanguage}. Nowadays, with the popularity of agent-based frameworks, researchers have begun to develop agent-based methods to solve complex tasks, such as OpenHands \cite{openhands}, AutoCodeRover \cite{autocoderover}, and SWE-Agent \cite{swe-agent}.

\vspace{-0.3ex}
\subsection{Automatic Compilation}
\vspace{-0.3ex}

In modern software development, there are a large number of open-source code repositories, but due to differences in project management and document writing among developers, the quality and standardization of compilation guides vary. Many projects lack detailed compilation instructions, which may cause users to encounter problems such as inconsistent environment configuration or lack of necessary dependencies when trying to compile. In addition, some open-source projects store compilation guides in external documents or websites without clearly marking them in the codebase, resulting in the compilation process that relies on manual steps, which is both error-prone and time-consuming. These problems make it more challenging to automate the compilation of open-source projects, and also highlight the importance of automated compilation tools in improving the maintainability and scalability of open-source projects.

Oss-Fuzz-Gen\cite{Liu_OSS-Fuzz-Gen_Automated_Fuzz_2024} is an open-source tool designed to fuzz real-world projects, including a part for building projects. This part works by analyzing the structure of the code repository and searching for specific files. Based on the presence of these files, a set of predefined compilation instructions is then executed to build the project. For example, if the repository contains \texttt{bootstrap.sh and Makefile.am}, Oss-Fuzz-Gen will execute the \texttt{\textquotedblleft./bootstrap.sh; ./configure; make\textquotedblright} commands in sequence to build the project. However, Oss-Fuzz-Gen may not be sufficient for projects where the specified files are absent. Additionally, the tool lacks adaptability to changing environments, making it less flexible in dynamic or evolving software projects. 

To be closer to realistic compilation scenarios, we formalize repo-level compilation tasks and propose CompileAgent to help LLMs complete this complex task. We also built a repo-level compilation benchmark CompileAgentBench to evaluate our approach and provide details of the benchmark in Appendix \ref{sec:benchdetails}. Compared with Oss-Fuzz-Gen, CompileAgent is more suitable for handling real-world compilation tasks.

\vspace{-0.5ex}
\section{Repo-Level Compilation Task}
\vspace{-0.5ex}

To bridge the gap between current compilation tasks and real-world software building scenarios, we formalized the repo-level compilation task. Unlike simple file-level compilation, code repositories often entail complex build configurations and interdependencies across multiple files. Consequently, an automated compile system as shown in Figure \ref{fig:automated compilation system}, which is an integrated tool or a comprehensive framework designed to facilitate the entire compilation process, must comprehend the entire repository, its dependencies, and the interactions between its components to ensure successful compilation at the repo-level. The repo-level compilation task focuses on managing the compilation process by considering all relevant software artifacts within the repository, including documentation, repository structure, and interactive environment.

\noindent\textbf{Documentation.} It provides essential insights into the project, including project introduction, configuration options, compilation guidelines, testing frameworks, and Demonstrations. Automated compile system can leverage it to extract and interpret information necessary for accurately configuring and executing the compilation process. Moreover, documentation often contains nuanced details about platform-specific dependencies or build settings that are critical for success.

\noindent\textbf{Repository Structure.} The structure of a repository reflects the organization and relationships among its files and modules. Effective repo-level compilation depends on a deep understanding these relationships, including dependency hierarchies between files or modules, and adhering to build sequence constraints(e.g., resolving \texttt{\textquotedblleft cmake\textquotedblright} configurations before invoking \texttt{\textquotedblleft make\textquotedblright}). Furthermore, addressing external library dependencies, such as linking with libraries like OpenSSL or Boost, is crucial for ensuring both compatibility and correctness. Efficiently navigating this structure is pivotal for repositories with intricate interdependencies.

\begin{figure*}[htbp]
  \centering
  \includegraphics[width=0.88\linewidth]{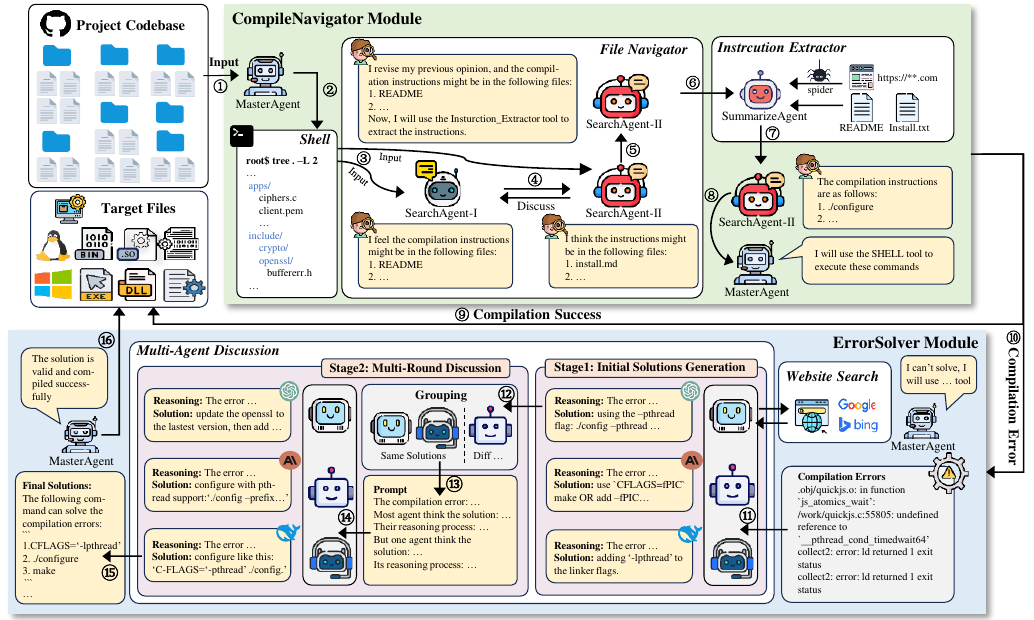}
  \vspace{-0.9ex}
  \caption{The overview of CompileAgent. By providing the repository of a given project, the automated compilation process can be seamlessly completed using the designed modules and agent strategy. Agents not explicitly specified are driven by DeepSeek-v2.5.}
  \vspace{-2.2ex}
  \label{fig:overview}
\end{figure*}

\noindent\textbf{Interactive Environment.} The interactive environment is integral to successful repo-level compilation, as it provides essential support throughout the process. It can provide detailed error messages and diagnostic information to the automated compile system during the compilation process, allowing it to identify and resolve issues in real time. This dynamic feedback loop allows the automated compile system to adjust the compilation process as needed, ensuring greater accuracy and efficiency. Additionally, the interactive environment should isolate the compilation process to safeguard the physical machine and provide independent build environments for each project.

In this paper, we consider LLM-based agent as an automated compilation system. Our objective is to rigorously evaluate its effectiveness in automating the repo-level compilation, ensuring that it can accurately identify the correct compilation instructions and efficiently resolve any issues that arise during the compilation process.

\vspace{-0.5ex}
\section{Method}
\vspace{-0.5ex}
In this section, we present the design of the LLM-based agent framework, CompileAgent, aimed at automating repo-level compilation. To effectively address the two key challenges mentioned in Section \ref{sec:Introduction}, we design two core modules, CompileNavigator and ErrorSolver, which together include five supporting tools, all integrated into a flow-based agent strategy, as shown in Figure \ref{fig:overview}.

\vspace{-0.3ex}
\subsection{Designed Module}
\vspace{-0.3ex}

When searching for compilation instructions in the given code repository, users typically rely on the repository's structure to identify potential files containing the necessary instructions. Moreover, when encountering difficulties during the compilation process that are hard to resolve, they often seek solutions through online resources, LLMs or other methods. To locate compilation instructions and resolve compilation errors, we model the process of solving the challenges and design the following two modules.

\subsubsection{CompileNavigator}

The CompileNavigator module is designed to tackle the challenge of finding the correct compilation instructions within a code repository. Typically, the necessary instructions are scattered across different documentation types, such as \texttt{README}, \texttt{doc.html}, \texttt{install.txt}, etc. making it difficult to locate them quickly. To address this challenge, the module employs three key tools: Shell, File Navigator, and Instruction Extractor. 

\noindent\textbf{Shell.} To ensure the security of physical machine during the compilation process, we isolate the entire compilation workflow from the host system by creating a container using Docker. The downloaded project is mounted into this container, and an SSH connection is established to access the terminal shell. The Docker container is built on the Ubuntu 22.04 operating system image. Through this tool, LLMs can interact with the interactive environment and execute any necessary commands.

\noindent\textbf{File Navigator.} To accurately locate the file containing the compilation instructions, we design two agents, SearchAgent \uppercase\expandafter{\romannumeral1} and SearchAgent \uppercase\expandafter{\romannumeral2}. The repository's structural information is provided as input, and the two agents engage in a collaborative discussion to determine the most likely file containing the compilation instructions.

\noindent\textbf{Instruction Extractor.} After identifying the files that likely contain the compilation instructions, the next task is to extract the instructions from them. In order to complete this, we design the SummarizeAgent, which reads the content of a specified file and searches for URLs related to compilation instructions within the file. If such URLs are found, requests are sent to retrieve the web page content. Finally, SummarizeAgent summarizes and outputs the relevant compilation instructions.

\subsubsection{ErrorSolver}

The ErrorSolver module is designed to address compilation errors during the project build process, which can stem from various issues such as syntax problems, missing dependencies, or configuration conflicts. To resolve these errors, we develop two key tools in this module: Website Search and Multi-Agent Discussion.

\noindent\textbf{Website Search.} Developers frequently publish solutions to compilation problems on websites, which search engines treat as valuable knowledge databases. When faced with similar problems, users can submit queries to search engines to find relevant solutions. Inspired by this, we encapsulate Google Search\footnote{https://www.google.com/} engine into a tool. However, since search results may include irrelevant content, we instruct the agents using the tool to prioritize reliable, open-source websites, like Github\footnote{https://github.com/} and StackOverflow\footnote{https://stackoverflow.com/}, and then aggregate the relevant information to provide a solution to the user's query. 

\noindent\textbf{Multi-Agent Discussion.} Although there are various single-agent approaches exist for solving reasoning tasks, such as self-polishing \cite{self-polish}, self-reflection \cite{self-reflection}, self-consistency \cite{self-consistency} and selection-inference \cite{selection-inference}, we think these complex reasoning approaches are unnecessary for solving compilation errors. Compilation errors typically come with clear error messages, such as path or environment configuration issues and compatibility problems. These errors can generally be resolved through straightforward analysis, consulting documentation, and making reasonable inferences, without the need of advanced reasoning processes. Inspired by Wang et al. \cite{rethinking-bounds} and reconcile \cite{reconcile}, we propose a Multi-Agent Discussion approach specifically designed to address compilation errors. In this method, multi-agents first analyze the complex compilation error and generate an initial solution. The agents then enter a multi-round discussion phase, where each can revise its analysis and response based on the inputs from the other agents in the previous round. The discussion continues until a consensus is reached or for up to R rounds. At the end of each round, the solutions, consisting of command lines, are segmented, and repeated terms are counted. If the number of repeated terms exceeds a specified threshold, the solutions are considered equivalent, and a final team response is generated. In this paper, we set up three agents for the discussion, with a maximum of 3 Rounds. 

\vspace{-0.3ex}
\subsection{Agent Strategy}
\vspace{-0.3ex}

When compiling a given project, users typically begin by consulting the project's compilation guide, and then execute the relevant compilation commands based on their environment. If issues arise during the process, they often resort to online searches or query tools like ChatGPT to troubleshoot until the compilation succeeds. Inspired by this workflow, to enable LLMs to effectively leverage our designed tools, we propose a flow-based agent strategy tailored for the automated compilation task.

The strategy defines the sequence in which tools are used and connects them seamlessly through prompts. MasterAgent is responsible for invoking the tools. The process is as follows:

\Circled{1} MasterAgent begins by downloading the target code repository to the local system and mounting it into the container using the Shell tool;

\Circled{2} Next, MasterAgent uses the Shell tool to run commands like \texttt{\textquotedblleft tree\textquotedblright} within the container to retrieve the repository structure;

\Circled{3} Then, MasterAgent invokes the FileNavigator tool to identify files that may contain the necessary compilation instructions;

\Circled{4} Subsequently, MasterAgent uses the InstructionExtractor tool to extract the compilation instructions and execute them via the Shell tool;

\Circled{5} If the Shell tool returns a successful compilation result, the compilation process is complete. If a compilation error occurs, MasterAgent first attempts to resolve the issue independently. If the issue persists after attempts, the ErrorSolver module is activated for several rounds of collaborative discussion. Finally, the compilation status is determined based on the Shell tool's outcome.

\vspace{-0.5ex}
\section{Experiment} \label{sec:experiment}
\vspace{-0.5ex}

We conduct extensive experiments to answer three research questions: (1) How much does CompileAgent improve the project compilation success rate compared to existing methods? (2) How effective is the flow-based strategy we designed when compared to existing agent strategies? (3) To what extent do the tools integrated within CompileAgent contribute to successful repo-level compilation?

\vspace{-0.3ex}
\subsection{Experimental Setup}
\vspace{-0.3ex}

\noindent\textbf{Benchmark.} To the best of our knowledge, there is no existing work that specifically evaluates repo-level compilation. Therefore, we manually construct a new benchmark for repo-level compilation to evaluate the effectiveness of our approach in this domain. We select 100 projects from many C/C++ projects on Github and carefully consider multiple factors during the project selection to ensure the authority and diversity of CompileAgentBench. First, we screen the projects based on the number of stars to ensure that the selected projects have high representativeness and practical value in the community. Moreover, we also consider the topics involved in the projects and finally select projects covering 14 different fields, including areas such as crypto, audio, and neural networks. On this basis, we also pay special attention to whether each project provided a clear compilation guide. Meanwhile, we arrange for three participants with 3 to 4 years of project development experience to manually compile these 100 projects to further verify the compilability of the selected projects and the accuracy of the evaluation. We finally obtain the target files of these 100 projects, and the entire compilation process took about 46 man-hours. More details refer to Appendix \ref{sec:benchdetails}. 

\noindent\textbf{Baselines.} As the first work dedicated to automating repo-level compilation, there is no related work for us to compare except Oss-Fuzz-Gen. However, there are some projects or technologies that are helpful for automated compilation tasks, such as the Readme-AI\footnote{https://github.com/eli64s/readme-ai} project and Retrival-Augumented Generation (RAG) techniques. 

Readme-AI is a developer tool that can generate well-structured and detailed documentation for a code repository based solely on its URL or file path. For cost-effectiveness, we utilize GPT-4o mini for documentation generation and specify in the requirements that the \texttt{\textquotedblleft How to compile/build from source code\textquotedblright} section should be included. A detailed example of this process is provided in Appendix \ref{sec:README-AI details}. 
RAG refers to a technique that enhances the output of LLMs by allowing them to reference external knowledge sources during response generation. In the compilation task, we leverage RAG as a tool. Specifically, we traverse the possible compilation files in the code repository, and then cut these file contents into chunks and generate vector embeddings. Each time the compilation instructions are searched for, LLMs generate instructions by retrieving the vector database. For a specific example, please refer to Appendix \ref{sec:RAG details}.

We also compare the flow-based agent strategy designed in this paper with existing agent strategies. According to the research of Wang et al. \cite{wang2024survey} and Xi et al. \cite{xi2023rise}, we select two common agent strategies that are suitable for the automated compilation task, including ReAct \cite{yao2022react}, Plan-and-Execute \cite{wang2023plan}. In addition, we also consider the comparison with OpenAIFunc \cite{OpenAIFunc}. 

\noindent\textbf{Base LLMs.} We apply CompileAgent to seven advanced LLMs, including three closed-source LLMs, i.e., GPT-4o \cite{gpt-4O}, Claude-3-5-sonnet \cite{claude-3-5-sonnet}, Gemini-1.5-flash \cite{gemini-1.5-flash}, as well as four open-source LLMs, i.e., Qwen2.5-32B-Instruct \cite{qwen2.5}, Mixtral-8×7B-Instruct \cite{mixtral-8×7B}, LLama3.1-70B-Instruct \cite{llama3.1-70B}, DeepSeek-v2.5 \cite{deepseekv2}. Additional descriptions are provided as a part of Table \ref{tab:table1}.

\noindent\textbf{Metrics.} In order to comprehensively evaluate the effectiveness of automated compilation tasks, we select three key indicators: compilation success rate, time cost, and expenses. Among these, the compilation success is determined when the target files in the precompiled projects completely match those generated by CompileAgent.

\begin{table*}[htbp]
    \centering
    \caption{The results of different baselines on CompileAgentBench.}
    \vspace{-1ex}
    \setlength{\tabcolsep}{0.96mm}
    \scalebox{0.78}{
    \begin{threeparttable}
    \begin{tabular}{l|c|ccccccccc|ccc}
        \toprule
        \multirow{2}{*}{\textbf{Models}} & \multirow{2}{*}{\textbf{Size}} & \multicolumn{3}{c}{\textbf{Oss-Fuzz-Gen\tnote{1}}} & \multicolumn{3}{c}{\textbf{Readme-AI}} & \multicolumn{3}{c|}{\textbf{RAG}} & \multicolumn{3}{c}{\textbf{CompileAgent}}\\
        \cmidrule(lr){3-5}\cmidrule(lr){6-8}\cmidrule(lr){9-11}\cmidrule(lr){12-14}
        & & $Csr\tnote{2}$ & $Time\tnote{3}$ & $Exp\tnote{4}$ & $Csr$ & $Time$ & $Exp$ & $Csr$ & $Time$ & $Exp$ & $Csr$ & $Time$ & $Exp$\\
        \midrule
        \textbf{\textit{Closed-source LLMs}} & & \multirow{5}{*}{25\%} & \multirow{5}{*}{53.01} & \multirow{5}{*}{-}  & & & &  & & & & &  \\
        GPT-4o \cite{gpt-4O} & - & & & & 72\% & 128.80 & 42.94 & 67\% & 11.12 & 45.78 & 89\% & 8.38 & 16.53 \\
        Claude-3-5-sonnet \cite{claude-3-5-sonnet} & - & & & & 79\% & 127.33 & 55.26 & 78\% & 8.30 & 54.44 & 96\% & 5.37 & 22.02 \\
        Gemini-1.5-flash \cite{gemini-1.5-flash} & - & & & & 41\% & 123.68 & 32.37 & 46\% & 9.28 & 35.72 & 65\% & 3.55 & 2.39 \\
        \midrule
        \textbf{\textit{Open-source LLMs}} & & \multirow{5}{*}{25\%} & \multirow{5}{*}{53.01} & \multirow{5}{*}{-} & & & &  & & & & & \\
        Qwen2.5-32B-Instruct \cite{qwen2.5} & 32B & & & & 70\% & 127.82 & 33.18 & 62\% & 10.55 & 36.73 & 80\% & 5.25 & 3.16 \\
        Mixtral-8×7B-Instruct \cite{mixtral-8×7B} & 42B & & & & 38\% & 124.60 & 33.12 & 45\% & 10.82 & 36.49 & 55\% & 4.88 & 4.32 \\
        LLama3.1-70B \cite{llama3.1-70B} & 70B & & & & 61\% & 125.03 & 33.57 & 61\% & 10.98 & 36.87 & 79\% & 7.38 & 2.71 \\
        DeepSeek-v2.5 \cite{deepseekv2} & 236B & & & & 71\% & 125.43 & 33.70 & 72\% & 11.30 & 36.08 & 91\% & 11.38 & 3.31 \\
        \bottomrule 
    \end{tabular}
    \begin{tablenotes}
        \footnotesize
        \item[1] The Oss-Fuzz-Gen project operates without relying on LLMs.
        \item[2] The proportion of successfully compiled projects to all projects.
        \item[3] The total duration required to complete the compilation process, measured in hours.
        \item[4] The total expense incurred during the compilation process, measured in US dollars.
    \end{tablenotes}
    \end{threeparttable}
    }
    \label{tab:table1}
\end{table*}

\begin{table*}[htbp]
    \centering
    \caption{The results of different agent strategies on CompileAgentBench.}
    \vspace{-1ex}
    \setlength{\tabcolsep}{0.96mm}
    \scalebox{0.78}{
    \begin{threeparttable}
    \begin{tabular}{l|c|ccccccccc|ccc}
        \toprule
        \multirow{2}{*}{\textbf{Models}} & \multirow{2}{*}{\textbf{Size}} & \multicolumn{3}{c}{\textbf{OpenAIFunc\tnote{1}}} & \multicolumn{3}{c}{\textbf{PlanAndExecute}} & \multicolumn{3}{c|}{\textbf{ReAct}} & \multicolumn{3}{c}{\textbf{Flow-based}}\\
        \cmidrule(lr){3-5}\cmidrule(lr){6-8}\cmidrule(lr){9-11}\cmidrule(lr){12-14}
        & & $Csr$ & $Time$ & $Exp$ & $Csr$ & $Time$ & $Exp$ & $Csr$ & $Time$ & $Exp$ & $Csr$ & $Time$ & $Exp$\\
        \midrule
        \textbf{\textit{Closed-source LLMs}} &  & & & & & & &  & & & & & \\
        GPT-4o \cite{gpt-4O} & - & 80\% & 6.75 & 22.51 & 40\% & 5.18 & 10.02 & 72\% & 6.58 & 23.63 & 89\% & 8.38 & 16.53 \\
        Claude-3-5-sonnet \cite{claude-3-5-sonnet} & - & - & - & - & 72\% & 5.02 & 13.77 & 81\% & 8.40 & 25.26 & 96\% & 5.37 & 22.02 \\
        \midrule
        \textbf{\textit{Open-source LLMs}} &  & & & & & & &  & & & & & \\
        LLama3.1-70B \cite{llama3.1-70B} & 70B & - & - & - & 26\% & 4.77 & 2.14 & 49\% & 10.48 & 6.52 & 79\% & 7.38 & 2.71 \\
        DeepSeek-v2.5 \cite{deepseekv2} & 236B & - & - & - & 70\% & 6.72 & 1.42 & 78\% & 11.32 & 3.88 & 91\% & 11.38 & 3.31 \\
        \bottomrule 
    \end{tabular}
    \begin{tablenotes}
        \footnotesize
        \item[1] The openaifunc refers to OpenAI's LLMs equipped with the capability to invoke functions.
    \end{tablenotes}
    \end{threeparttable}
    }
    \vspace{-2ex}
    \label{tab:table2}
\end{table*}

\vspace{-0.3ex}
\subsection{Repo-Level Compilation Performance}
\vspace{-0.3ex}

In this experiment, we use the specially designed repo-level benchmark, CompileAgentBench, to evaluate the performance of CompileAgent and three baselines in compiling code repositories across seven well-known LLMs. The results are presented in Table \ref{tab:table1}.

It turns out that our proposed CompileAgentBench is more challenging when not using LLMs methods, as evidenced by the lower compilation success rate of Oss-Fuzz-Gen. Compared with existing baselines, CompileAgent has significant performance improvements on LLMs with various sizes. Specifically, CompileAgent achieves the highest performance on the Claude-3-5-sonnet model, improving by 71\%, 17\%, and 18\% over all baselines, respectively; in terms of time cost, it saves 47.64 hours, 121.96 hours, and 2.93 hours; in terms of expenses, the average cost per project is only \$0.22. Excluding Oss-Fuzz-Gen, the total cost is reduced by \$33.24 and \$32.42, respectively. The performance improvement on other LLMs ranges from 30\% to 71\%, 10\% to 24\%, and 10\% to 22\%, which clearly demonstrates the effectiveness of our method. This indicates that the integrated tools in CompileAgent can effectively assist LLMs in completing the compilation process, meeting the real-world needs of repo-level compilation.

In addition, we also find that the more advanced LLMs tend to show better performance with CompileAgent. However, for the poor performance of Mixtral-8×7B-Instruct, we speculate that may be related to its model architecture design.

\vspace{-0.3ex}
\subsection{Strategy Performance}
\vspace{-0.3ex}

We also evaluate the impact of different agent strategies on CompileAgent, and make slight modifications to other strategies, enabling them to call the tool we designed. Additionally, we strategically select a set of representative LLMs for evaluation, considering the constraints of available resources and computing power. Table \ref{tab:table2} summarizes the experimental results of the evaluation. 

Our flow-based agent strategy achieves the highest compilation success rate on Claude-3-5-sonnet, but it also brings a lot of costs. It is worth noting that the success rate of each compilation strategy generally decreases when using LLMs with fewer parameters. Despite this, our designed strategy can still achieve a 30\%-53\% higher success rate than other agent strategies while maintaining low time and cost. These findings emphasize that the flow-based agent strategy we designed can also maintain a high compilation success rate even under LLMs with different parameter specifications, showing stronger robustness than other agent strategies.

Additionally, combined with the results of the first experiment, we find that the ReAct and Flow-based strategies are more suitable for the compilation task, and the PlanAndExecute strategy appears less suited for the task.

\vspace{-0.3ex}
\subsection{Ablation Study}
\vspace{-0.3ex}

In order to evaluate the impact of our designed tools on CompileAgent, we conduct an ablation study. In this experiment, we select GPT-4o with Flow-based as the ablation subject and record the usage frequency of each tool during the compilation process. We then perform the ablation of these tools, and the results are presented in Table \ref{tab:tabel3}.

Our experimental results indicate that the Multi-Agent Discussion tool is the most frequently called in the compilation task. Ablating this tool leads to a significant drop in the compilation success rate, reaching 18\%, while the time and cost overhead required for compilation also increase. This suggests that CompileAgent relies heavily on the tool when tackling complex problems, as it plays a crucial role in enhancing both accuracy and efficiency. Moreover, the ablation results of the other tools demonstrate their positive contributions to the performance of CompileAgent to varying degrees. Overall, the ablation experiment results confirm the effectiveness and practicality of the tools we designed for real-world compilation tasks.

\begin{table}[t]
    \centering
    \caption{Average tool usage number and ablation result on CompileAgentBench for CompileAgent which is based on GPT-4o.}
    \vspace{-1.5ex}
    \setlength{\tabcolsep}{1.2mm}
    \scalebox{0.86}{
    \begin{threeparttable}
    \begin{tabular}{l|c|ccc}
        \toprule
         \multirow{2}{*}{\textbf{Tools}} & \multirow{2}{*}{\textbf{Usage}} & \multicolumn{3}{c}{\textbf{Ablation Result}}\\
         \cmidrule(lr){3-5}
         & & $Csr$ & $Time$ & $Exp$ \\
         \midrule
         \textbf{\textit{CompileAgent}} & - & 89\% & 8.38 & 16.53 \\
         \midrule
         \emph{Shell\tnote{1}} & - & - & - & - \\
         \emph{File Navigator} & 1.21 & 81\% & 6.93 & 17.32 \\
         \emph{Instruction Extractor\tnote{2}} & 1.63 & 77\% & 7.18 & 18.26 \\
         \emph{Website Search} & 0.61 & 84\% & 7.25 & 16.53 \\
         \emph{Multi-Agent Discussion} & 1.87 & 71\% & 8.77 & 18.89\\
        \bottomrule 
    \end{tabular}
    \begin{tablenotes}
        \footnotesize
        \item[1] The Shell tool is essential for executing compilation instructions and is a necessary condition for compilation tasks.
        \item[2] We retain the core functionality of the Instruction Extractor while removing the web content crawling feature.
    \end{tablenotes}
    \end{threeparttable}
    }
    \vspace{-1.8ex}
    \label{tab:tabel3}
\end{table}

\section{Discussion}
\vspace{-0.5ex}

\vspace{-0.3ex}
\subsection{Failure Analysis}
\vspace{-0.3ex}

In the previous experiments, CompileAgent encounters several compilation failures. After analysis, we summarize the most common three errors in the compilation process: \uppercase\expandafter{\romannumeral1)} Complex Build Dependencies. Some projects rely on intricate dependency chains involving specific versions of libraries, and missing or incompatible dependencies lead to building failures. \uppercase\expandafter{\romannumeral2)} Toolchain Mismatch. Some projects require specific versions of compilers, interpreters, or build tools that are not available or configured properly in the CompileAgent environment, resulting in compilation errors. \uppercase\expandafter{\romannumeral3)} Configuration Complexity. The complex configuration settings in some projects, such as unmatched environmental variables and improperly defined parameters, resulting in the failure of compilation.

\vspace{-0.3ex}
\subsection{Multi-Language and Multi-Architecture Compilation}
\vspace{-0.3ex}

Although the CompileAgent in this article is mainly designed for C/C++ projects, it can also support multi-language and multi-architecture compilation due to its scalability and flexibility, and can be expanded to realize the automated compilation process in different environments.

For multi-language compilation, we can first install the interactive environment of each language in Docker and dynamically adjust the toolchain by detecting the programming language used by the project. This includes selecting the appropriate compiler and configuring language-specific build tools, such as javac for Java or npm for JavaScript.

For multi-architecture compilation, we can use the system emulation tools provided by QEMU\footnote{https://www.qemu.org/} to enable CompileAgent to interact with environments of different processor architectures such as ARM, MIPS, and X86 to achieve cross-platform compilation.

\vspace{-0.3ex}
\subsection{Large-Scale Code Analysis}
\vspace{-0.3ex}

By integrating with multiple code analysis tools, CompileAgent can evaluate the security of repositories during the compilation process, further ensuring the reliability of compilation results, especially for some potentially malicious code repositories. Specifically, we can encapsulate tools such as Coverity Scan\footnote{https://scan.coverity.com/} and the Scan-Build\footnote{https://github.com/llvm/llvm-project} and call them to perform security analysis when CompileAgent performs compilation, identifying critical vulnerabilities, including buffer overflows or unsafe practices.

\vspace{-0.5ex}
\section{Conclusion}
\vspace{-0.5ex}

In this paper, we propose CompileAgent, the first LLM-based agent framework designed for repo-level compilation, which integrates five tools and a flow-based agent strategy to enable LLMs to interact with software artifacts. To assess its performance, we construct a public repo-level compilation benchmark CompileAgentBench, and establish two compilation-friendly schemes as baselines. Experimental results on multiple LLMs demonstrate the effectiveness of CompileAgent. Finally, We also highlight the scalability of CompileAgent and expand its application prospects.

\section*{Limitations}

Our work is the first attempt to use LLM-based agents to handle the repo-level compilation task, and verify the effectiveness of CompileAgent through comprehensive experiments. However, there are still some limitations that need to be further addressed in the future:

Firstly, CompileAgent relies on the understanding capability of LLMs. During compilation, the agents may misinterpret prompts or instructions, leading to repeated or incorrect actions, which impacts its efficiency in resolving compilation issues. Future work will explore fine-tuning models to improve their in interpreting instructions.

Secondly, the tools incorporated into CompileAgent are relatively basic, leaving unexplored potential for leveraging more advanced programming and debugging tools. Later we can expand the toolset to improve the performance of agents in tackling intricate compilation tasks and error resolution.

Finally, since CompileAgent is highly dependent on the quality of prompt engineering, optimizing the prompts used in the agent system is crucial for its performance. In the future work, we will explore more effective agent strategies to improve overall system performance.

\section*{Ethics Consideration}

We promise that CompileAgent is inspired by real-world needs for code repositories compilation, with CompileAgentBench constructed from real-world code repositories to ensure practical relevance. During our experiments, all projects were manually reviewed to verify the absence of private information or offensive content. Additionally, we manually compiled each project to validate the reliability of CompileAgentBench. 

\bibliography{main}

\begin{thebibliography}{40}
\providecommand{\natexlab}[1]{#1}

\bibitem[{Bianou and Batogna(2024)}]{pentest-ai}
Stanislas~G. Bianou and Rodrigue~G. Batogna. 2024.
\newblock \href {https://doi.org/10.1109/CSR61664.2024.10679480} {Pentest-ai,
  an llm-powered multi-agents framework for penetration testing automation
  leveraging mitre attack}.
\newblock In \emph{2024 IEEE International Conference on Cyber Security and
  Resilience (CSR)}, pages 763--770.

\bibitem[{Bouzenia et~al.(2024)Bouzenia, Devanbu, and Pradel}]{repairagent}
Islem Bouzenia, Premkumar Devanbu, and Michael Pradel. 2024.
\newblock Repairagent: An autonomous, llm-based agent for program repair.
\newblock \emph{arXiv preprint arXiv:2403.17134}.

\bibitem[{Chen et~al.(2024)Chen, Saha, and Bansal}]{reconcile}
Justin Chen, Swarnadeep Saha, and Mohit Bansal. 2024.
\newblock \href {https://doi.org/10.18653/v1/2024.acl-long.381} {{R}e{C}oncile:
  Round-table conference improves reasoning via consensus among diverse
  {LLM}s}.
\newblock In \emph{Proceedings of the 62nd Annual Meeting of the Association
  for Computational Linguistics (Volume 1: Long Papers)}, pages 7066--7085,
  Bangkok, Thailand. Association for Computational Linguistics.

\bibitem[{Chen et~al.(2023)Chen, Zhang, Ren, Zhao, Cai, Wang, Wang, Liu, and
  Chang}]{chen2023towards}
Liang Chen, Yichi Zhang, Shuhuai Ren, Haozhe Zhao, Zefan Cai, Yuchi Wang, Peiyi
  Wang, Tianyu Liu, and Baobao Chang. 2023.
\newblock Towards end-to-end embodied decision making via multi-modal large
  language model: Explorations with gpt4-vision and beyond.
\newblock \emph{arXiv preprint arXiv:2310.02071}.

\bibitem[{Claude(2024)}]{claude-3-5-sonnet}
Claude. 2024.
\newblock \href {https://www.anthropic.com/claude/sonnet}
  {\url{https://www.anthropic.com/claude/sonnet}}.

\bibitem[{Creswell et~al.(2022)Creswell, Shanahan, and
  Higgins}]{selection-inference}
Antonia Creswell, Murray Shanahan, and Irina Higgins. 2022.
\newblock \href {https://arxiv.org/abs/2205.09712} {Selection-inference:
  Exploiting large language models for interpretable logical reasoning}.
\newblock \emph{Preprint}, arXiv:2205.09712.

\bibitem[{DeepSeek-AI(2024)}]{deepseekv2}
DeepSeek-AI. 2024.
\newblock \href {https://arxiv.org/abs/2405.04434} {Deepseek-v2: A strong,
  economical, and efficient mixture-of-experts language model}.
\newblock \emph{Preprint}, arXiv:2405.04434.

\bibitem[{Deng et~al.(2024)Deng, Liu, Mayoral-Vilches, Liu, Li, Xu, Zhang, Liu,
  Pinzger, and Rass}]{pentestgpt}
Gelei Deng, Yi~Liu, V{\'\i}ctor Mayoral-Vilches, Peng Liu, Yuekang Li, Yuan Xu,
  Tianwei Zhang, Yang Liu, Martin Pinzger, and Stefan Rass. 2024.
\newblock $\{$PentestGPT$\}$: Evaluating and harnessing large language models
  for automated penetration testing.
\newblock In \emph{33rd USENIX Security Symposium (USENIX Security 24)}, pages
  847--864.

\bibitem[{Gemini(2024)}]{gemini-1.5-flash}
Gemini. 2024.
\newblock \href {https://deepmind.google/technologies/gemini/flash}
  {\url{https://deepmind.google/technologies/gemini/flash}}.

\bibitem[{GPT-4o(2024)}]{gpt-4O}
GPT-4o. 2024.
\newblock \href {https://platform.openai.com/docs/models/gpt-4o}
  {\url{https://platform.openai.com/docs/models/gpt-4o}}.

\bibitem[{Huang et~al.(2023)Huang, Bu, Zhang, Luck, and Cui}]{agentcoder}
Dong Huang, Qingwen Bu, Jie~M Zhang, Michael Luck, and Heming Cui. 2023.
\newblock Agentcoder: Multi-agent-based code generation with iterative testing
  and optimisation.
\newblock \emph{arXiv preprint arXiv:2312.13010}.

\bibitem[{Huang et~al.(2024)Huang, Cheng, Huang, Shen, Xu, Zhang, and
  Qu}]{queryagent}
Xiang Huang, Sitao Cheng, Shanshan Huang, Jiayu Shen, Yong Xu, Chaoyun Zhang,
  and Yuzhong Qu. 2024.
\newblock \href {https://doi.org/10.18653/v1/2024.acl-long.274}
  {{Q}uery{A}gent: A reliable and efficient reasoning framework with
  environmental feedback based self-correction}.
\newblock In \emph{Proceedings of the 62nd Annual Meeting of the Association
  for Computational Linguistics (Volume 1: Long Papers)}, pages 5014--5035,
  Bangkok, Thailand. Association for Computational Linguistics.

\bibitem[{Islam et~al.(2024)Islam, Ali, and Parvez}]{mapcoder}
Md.~Ashraful Islam, Mohammed~Eunus Ali, and Md~Rizwan Parvez. 2024.
\newblock \href {https://doi.org/10.18653/v1/2024.acl-long.269} {{M}ap{C}oder:
  Multi-agent code generation for competitive problem solving}.
\newblock In \emph{Proceedings of the 62nd Annual Meeting of the Association
  for Computational Linguistics (Volume 1: Long Papers)}, pages 4912--4944,
  Bangkok, Thailand. Association for Computational Linguistics.

\bibitem[{Jiang et~al.(2024)Jiang, An, Huang, Tang, Nie, Wu, and
  Zhang}]{BinaryAI}
Ling Jiang, Junwen An, Huihui Huang, Qiyi Tang, Sen Nie, Shi Wu, and Yuqun
  Zhang. 2024.
\newblock \href {https://doi.org/10.1145/3597503.3639100} {Binaryai: Binary
  software composition analysis via intelligent binary source code matching}.
\newblock In \emph{Proceedings of the IEEE/ACM 46th International Conference on
  Software Engineering}, ICSE '24, New York, NY, USA. Association for Computing
  Machinery.

\bibitem[{Liu et~al.(2024{\natexlab{a}})Liu, Chang, metzman, Sablotny, and
  Maruseac}]{Liu_OSS-Fuzz-Gen_Automated_Fuzz_2024}
Dongge Liu, Oliver Chang, Jonathan metzman, Martin Sablotny, and Mihai
  Maruseac. 2024{\natexlab{a}}.
\newblock \href {https://github.com/google/oss-fuzz-gen} {{OSS-Fuzz-Gen:
  Automated Fuzz Target Generation}}.

\bibitem[{Liu et~al.(2024{\natexlab{b}})Liu, Gao, Wang, Liu, Shi, Zhang, and
  Peng}]{marscode}
Yizhou Liu, Pengfei Gao, Xinchen Wang, Jie Liu, Yexuan Shi, Zhao Zhang, and
  Chao Peng. 2024{\natexlab{b}}.
\newblock Marscode agent: Ai-native automated bug fixing.
\newblock \emph{arXiv preprint arXiv:2409.00899}.

\bibitem[{Meta-LLaMa(2024)}]{llama3.1-70B}
Meta-LLaMa. 2024.
\newblock \href {https://huggingface.co/meta-llama/Llama-3.1-70B}
  {\url{https://huggingface.co/meta-llama/Llama-3.1-70B}}.

\bibitem[{MistralAI(2023)}]{mixtral-8×7B}
MistralAI. 2023.
\newblock \href {https://huggingface.co/mistralai/Mixtral-8x7B-Instruct-v0.1}
  {\url{https://huggingface.co/mistralai/Mixtral-8x7B-Instruct-v0.1}}.

\bibitem[{OpenAI(2023)}]{OpenAIFunc}
OpenAI. 2023.
\newblock \href
  {https://openai.com/index/function-calling-and-other-api-updates/}
  {\url{https://openai.com/index/function-calling-and-other-api-updates/}}.

\bibitem[{OpenAI(2024)}]{text-embedding-3-large}
OpenAI. 2024.
\newblock \href
  {https://openai.com/index/new-embedding-models-and-api-updates/}
  {\url{https://openai.com/index/new-embedding-models-and-api-updates/}}.

\bibitem[{Qian et~al.(2024{\natexlab{a}})Qian, Liu, Liu, Chen, Dang, Li, Yang,
  Chen, Su, Cong, Xu, Li, Liu, and Sun}]{chatdev}
Chen Qian, Wei Liu, Hongzhang Liu, Nuo Chen, Yufan Dang, Jiahao Li, Cheng Yang,
  Weize Chen, Yusheng Su, Xin Cong, Juyuan Xu, Dahai Li, Zhiyuan Liu, and
  Maosong Sun. 2024{\natexlab{a}}.
\newblock \href {https://doi.org/10.18653/v1/2024.acl-long.810} {{C}hat{D}ev:
  Communicative agents for software development}.
\newblock In \emph{Proceedings of the 62nd Annual Meeting of the Association
  for Computational Linguistics (Volume 1: Long Papers)}, pages 15174--15186,
  Bangkok, Thailand. Association for Computational Linguistics.

\bibitem[{Qian et~al.(2024{\natexlab{b}})Qian, He, Zhuang, Deng, Qin, Cong,
  Zhang, Zhou, Lin, Liu, and Sun}]{tellmemore}
Cheng Qian, Bingxiang He, Zhong Zhuang, Jia Deng, Yujia Qin, Xin Cong, Zhong
  Zhang, Jie Zhou, Yankai Lin, Zhiyuan Liu, and Maosong Sun.
  2024{\natexlab{b}}.
\newblock \href {https://doi.org/10.18653/v1/2024.acl-long.61} {Tell me more!
  towards implicit user intention understanding of language model driven
  agents}.
\newblock In \emph{Proceedings of the 62nd Annual Meeting of the Association
  for Computational Linguistics (Volume 1: Long Papers)}, pages 1088--1113,
  Bangkok, Thailand. Association for Computational Linguistics.

\bibitem[{Shen et~al.(2024)Shen, Wang, Li, Chen, Zhao, Sun, Wang, and
  Ruan}]{pentestagent}
Xiangmin Shen, Lingzhi Wang, Zhenyuan Li, Yan Chen, Wencheng Zhao, Dawei Sun,
  Jiashui Wang, and Wei Ruan. 2024.
\newblock Pentestagent: Incorporating llm agents to automated penetration
  testing.
\newblock \emph{arXiv preprint arXiv:2411.05185}.

\bibitem[{Tan et~al.(2020)Tan, Xie, Li, Barker, and Tumeo}]{OpenCGRA}
Cheng Tan, Chenhao Xie, Ang Li, Kevin~J. Barker, and Antonino Tumeo. 2020.
\newblock \href {https://doi.org/10.1109/ICCD50377.2020.00070} {Opencgra: An
  open-source unified framework for modeling, testing, and evaluating cgras}.
\newblock In \emph{2020 IEEE 38th International Conference on Computer Design
  (ICCD)}, pages 381--388.

\bibitem[{Team(2024)}]{qwen2.5}
Qwen Team. 2024.
\newblock \href {https://qwenlm.github.io/blog/qwen2.5/} {Qwen2.5: A party of
  foundation models}.

\bibitem[{Wang et~al.(2024{\natexlab{a}})Wang, Prasad, Stengel-Eskin, and
  Bansal}]{self-consistency}
Han Wang, Archiki Prasad, Elias Stengel-Eskin, and Mohit Bansal.
  2024{\natexlab{a}}.
\newblock \href {https://doi.org/10.18653/v1/2024.acl-short.28} {Soft
  self-consistency improves language models agents}.
\newblock In \emph{Proceedings of the 62nd Annual Meeting of the Association
  for Computational Linguistics (Volume 2: Short Papers)}, pages 287--301,
  Bangkok, Thailand. Association for Computational Linguistics.

\bibitem[{Wang et~al.(2024{\natexlab{b}})Wang, Gao, Zhang, Sha, Sun, Zhou, Zhu,
  Sun, Qiu, and Xiao}]{CLAP}
Hao Wang, Zeyu Gao, Chao Zhang, Zihan Sha, Mingyang Sun, Yuchen Zhou, Wenyu
  Zhu, Wenju Sun, Han Qiu, and Xi~Xiao. 2024{\natexlab{b}}.
\newblock \href {https://doi.org/10.1145/3650212.3652145} {Clap: Learning
  transferable binary code representations with natural language supervision}.
\newblock In \emph{Proceedings of the 33rd ACM SIGSOFT International Symposium
  on Software Testing and Analysis}, ISSTA 2024, page 503–515, New York, NY,
  USA. Association for Computing Machinery.

\bibitem[{Wang et~al.(2024{\natexlab{c}})Wang, Ma, Feng, Zhang, Yang, Zhang,
  Chen, Tang, Chen, Lin et~al.}]{wang2024survey}
Lei Wang, Chen Ma, Xueyang Feng, Zeyu Zhang, Hao Yang, Jingsen Zhang, Zhiyuan
  Chen, Jiakai Tang, Xu~Chen, Yankai Lin, et~al. 2024{\natexlab{c}}.
\newblock A survey on large language model based autonomous agents.
\newblock \emph{Frontiers of Computer Science}, 18(6):186345.

\bibitem[{Wang et~al.(2023)Wang, Xu, Lan, Hu, Lan, Lee, and Lim}]{wang2023plan}
Lei Wang, Wanyu Xu, Yihuai Lan, Zhiqiang Hu, Yunshi Lan, Roy Ka-Wei Lee, and
  Ee-Peng Lim. 2023.
\newblock Plan-and-solve prompting: Improving zero-shot chain-of-thought
  reasoning by large language models.
\newblock \emph{arXiv preprint arXiv:2305.04091}.

\bibitem[{Wang et~al.(2024{\natexlab{d}})Wang, Wang, Su, Tong, and
  Song}]{rethinking-bounds}
Qineng Wang, Zihao Wang, Ying Su, Hanghang Tong, and Yangqiu Song.
  2024{\natexlab{d}}.
\newblock \href {https://doi.org/10.18653/v1/2024.acl-long.331} {Rethinking the
  bounds of {LLM} reasoning: Are multi-agent discussions the key?}
\newblock In \emph{Proceedings of the 62nd Annual Meeting of the Association
  for Computational Linguistics (Volume 1: Long Papers)}, pages 6106--6131,
  Bangkok, Thailand. Association for Computational Linguistics.

\bibitem[{Wang et~al.(2024{\natexlab{e}})Wang, Li, Song, Xu, Tang, Zhuge, Pan,
  Song, Li, Singh, Tran, Li, Ma, Zheng, Qian, Shao, Muennighoff, Zhang, Hui,
  Lin, Brennan, Peng, Ji, and Neubig}]{openhands}
Xingyao Wang, Boxuan Li, Yufan Song, Frank~F. Xu, Xiangru Tang, Mingchen Zhuge,
  Jiayi Pan, Yueqi Song, Bowen Li, Jaskirat Singh, Hoang~H. Tran, Fuqiang Li,
  Ren Ma, Mingzhang Zheng, Bill Qian, Yanjun Shao, Niklas Muennighoff, Yizhe
  Zhang, Binyuan Hui, Junyang Lin, Robert Brennan, Hao Peng, Heng Ji, and
  Graham Neubig. 2024{\natexlab{e}}.
\newblock \href {https://arxiv.org/abs/2407.16741} {{OpenHands: An Open
  Platform for AI Software Developers as Generalist Agents}}.
\newblock \emph{Preprint}, arXiv:2407.16741.

\bibitem[{Xi et~al.(2023{\natexlab{a}})Xi, Chen, Guo, He, Ding, Hong, Zhang,
  Wang, Jin, Zhou et~al.}]{xi2023rise}
Zhiheng Xi, Wenxiang Chen, Xin Guo, Wei He, Yiwen Ding, Boyang Hong, Ming
  Zhang, Junzhe Wang, Senjie Jin, Enyu Zhou, et~al. 2023{\natexlab{a}}.
\newblock The rise and potential of large language model based agents: A
  survey.
\newblock \emph{arXiv preprint arXiv:2309.07864}.

\bibitem[{Xi et~al.(2023{\natexlab{b}})Xi, Jin, Zhou, Zheng, Gao, Liu, Gui,
  Zhang, and Huang}]{self-polish}
Zhiheng Xi, Senjie Jin, Yuhao Zhou, Rui Zheng, Songyang Gao, Jia Liu, Tao Gui,
  Qi~Zhang, and Xuanjing Huang. 2023{\natexlab{b}}.
\newblock \href {https://doi.org/10.18653/v1/2023.findings-emnlp.762}
  {Self-{P}olish: Enhance reasoning in large language models via problem
  refinement}.
\newblock In \emph{Findings of the Association for Computational Linguistics:
  EMNLP 2023}, pages 11383--11406, Singapore. Association for Computational
  Linguistics.

\bibitem[{Xie et~al.(2023)Xie, Zhou, Cheng, Shi, Weng, Liu, Hua, Zhao, Liu,
  Liu, Liu, Xu, Su, Shin, Xiong, and
  Yu}]{xie2023openagentsopenplatformlanguage}
Tianbao Xie, Fan Zhou, Zhoujun Cheng, Peng Shi, Luoxuan Weng, Yitao Liu,
  Toh~Jing Hua, Junning Zhao, Qian Liu, Che Liu, Leo~Z. Liu, Yiheng Xu, Hongjin
  Su, Dongchan Shin, Caiming Xiong, and Tao Yu. 2023.
\newblock \href {https://arxiv.org/abs/2310.10634} {Openagents: An open
  platform for language agents in the wild}.
\newblock \emph{Preprint}, arXiv:2310.10634.

\bibitem[{Yan et~al.(2024)Yan, Zhu, Wang, Gui, and He}]{self-reflection}
Hanqi Yan, Qinglin Zhu, Xinyu Wang, Lin Gui, and Yulan He. 2024.
\newblock \href {https://doi.org/10.18653/v1/2024.acl-long.382} {Mirror:
  Multiple-perspective self-reflection method for knowledge-rich reasoning}.
\newblock In \emph{Proceedings of the 62nd Annual Meeting of the Association
  for Computational Linguistics (Volume 1: Long Papers)}, pages 7086--7103,
  Bangkok, Thailand. Association for Computational Linguistics.

\bibitem[{Yang et~al.(2024)Yang, Jimenez, Wettig, Lieret, Yao, Narasimhan, and
  Press}]{swe-agent}
John Yang, Carlos~E. Jimenez, Alexander Wettig, Kilian Lieret, Shunyu Yao,
  Karthik Narasimhan, and Ofir Press. 2024.
\newblock \href {https://arxiv.org/abs/2405.15793} {Swe-agent: Agent-computer
  interfaces enable automated software engineering}.
\newblock \emph{Preprint}, arXiv:2405.15793.

\bibitem[{Yao et~al.(2022)Yao, Zhao, Yu, Du, Shafran, Narasimhan, and
  Cao}]{yao2022react}
Shunyu Yao, Jeffrey Zhao, Dian Yu, Nan Du, Izhak Shafran, Karthik Narasimhan,
  and Yuan Cao. 2022.
\newblock React: Synergizing reasoning and acting in language models.
\newblock \emph{arXiv preprint arXiv:2210.03629}.

\bibitem[{Ye et~al.(2023)Ye, Wu, Ma, Zhang, Du, Liu, Ji, and Wang}]{CP-BCS}
Tong Ye, Lingfei Wu, Tengfei Ma, Xuhong Zhang, Yangkai Du, Peiyu Liu, Shouling
  Ji, and Wenhai Wang. 2023.
\newblock \href {https://doi.org/10.18653/v1/2023.emnlp-main.911} {{CP}-{BCS}:
  Binary code summarization guided by control flow graph and pseudo code}.
\newblock In \emph{Proceedings of the 2023 Conference on Empirical Methods in
  Natural Language Processing}, pages 14740--14752, Singapore. Association for
  Computational Linguistics.

\bibitem[{Zhang et~al.(2024{\natexlab{a}})Zhang, Li, Li, Shi, and
  Jin}]{CodeAgent}
Kechi Zhang, Jia Li, Ge~Li, Xianjie Shi, and Zhi Jin. 2024{\natexlab{a}}.
\newblock \href {https://doi.org/10.18653/v1/2024.acl-long.737} {{C}ode{A}gent:
  Enhancing code generation with tool-integrated agent systems for real-world
  repo-level coding challenges}.
\newblock In \emph{Proceedings of the 62nd Annual Meeting of the Association
  for Computational Linguistics (Volume 1: Long Papers)}, pages 13643--13658,
  Bangkok, Thailand. Association for Computational Linguistics.

\bibitem[{Zhang et~al.(2024{\natexlab{b}})Zhang, Ruan, Fan, and
  Roychoudhury}]{autocoderover}
Yuntong Zhang, Haifeng Ruan, Zhiyu Fan, and Abhik Roychoudhury.
  2024{\natexlab{b}}.
\newblock \href {https://doi.org/10.1145/3650212.3680384} {Autocoderover:
  Autonomous program improvement}.
\newblock In \emph{Proceedings of the 33rd ACM SIGSOFT International Symposium
  on Software Testing and Analysis}, ISSTA 2024, page 1592–1604, New York,
  NY, USA. Association for Computing Machinery.

\end{thebibliography}

\appendix

\section{Benchmark Details}
\label{sec:benchdetails}

Table \ref{tab:table4} presents the composition of CompileAgentBench, which includes 100 popular projects across 14 topics. To align with the distribution of compilation guides in real-world code repositories, CompileAgentBench maintains a ratio of compilation guides in repo to those not in repo, as well as those without guides, at 7:2:1.

\section{Readme-AI Details}
\label{sec:README-AI details}

\begin{figure}[htbp]
  \centering
  \includegraphics[width=0.95\linewidth]{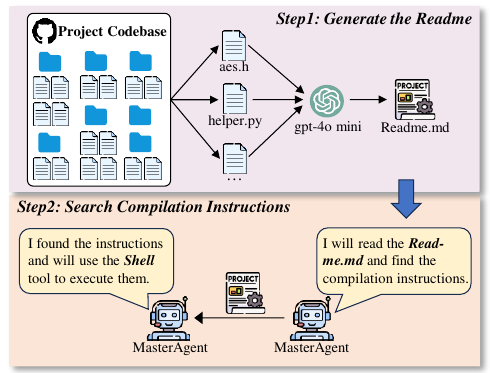}
  \caption{The details of Readme-AI.}
  \label{fig:readme-ai}
\end{figure}

Figure \ref{fig:readme-ai} shows the Readme-AI how to be used in our compilation task. Its workflow is that GPT-4o mini first traverses all project files, generate a \texttt{Readme.md} file based on specific requirements, and finally MasterAgent can find the compilation instructions by reading the \texttt{Readme.md}.

\section{RAG Details}
\label{sec:RAG details}

\begin{figure}[htbp]
  \centering
  \includegraphics[width=0.95\linewidth]{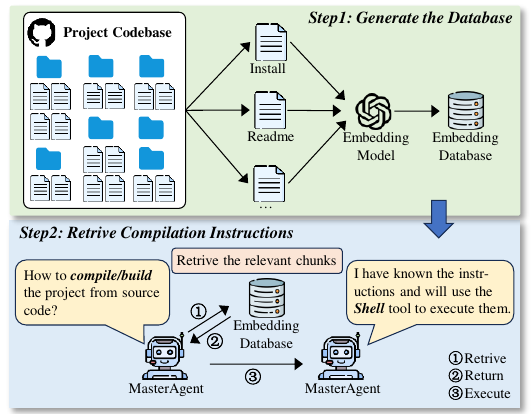}
  \caption{The details of RAG.}
  \label{fig:rag}
\end{figure}

Figure \ref{fig:rag} illustrates how the RAG technology is applied in our compilation task. We first specify some files that may contain compilation instructions, such as README, INSTALL, etc., and then split the contents of the files into chunks and generate embeddings and store them in the embedding database. Finally, MasterAgent retrives the embedding database to obtain the compilation instructions. The embedding model we use in this article is text-embedding-3-large \cite{text-embedding-3-large}.

\begin{table*}[htbp]
    \centering
    \caption{The composition of  CompileAgentBench.}
    \setlength{\tabcolsep}{0.90mm}
    \scalebox{0.78}{
    \begin{threeparttable}
    \begin{tabular}{l|c|cc|c||l|c|cc|c}
        \toprule
        \multirow{2}{*}{\textbf{Project}} & \multirow{2}{*}{\textbf{Topic}} & \multicolumn{2}{c|}{\textbf{Existing Guide}} & \multirow{2}{*}{\textbf{No Guide}} & \multirow{2}{*}{\textbf{Project}} & \multirow{2}{*}{\textbf{Topic}} & \multicolumn{2}{c|}{\textbf{Existing Guide}} & \multirow{2}{*}{\textbf{No Guide}} \\
        \cmidrule(lr){3-4}\cmidrule(lr){8-9}
        & & $In Repo$ & $Not In Repo$ & & & & $In Repo$ & $Not In Repo$ & \\
        \midrule
        FFmpeg & Audio & \checkmark & \error & \error & libvips & Image & \checkmark & \error & \error \\
        aubio & Audio & \checkmark & \error & \error & mozjpeg & Image & \checkmark & \error & \error \\
        cava & Audio & \checkmark & \error & \error & clib & Linux & \checkmark & \error & \error \\
        Julius & Audio & \checkmark & \error & \error & activate-linux & Linux & \checkmark & \error & \error \\
        zstd & Compression & \checkmark & \error & \error & libbpf & Linux & \checkmark & \error & \error \\
        7z & Compression & \error & \checkmark & \error & util-linux & Linux & \checkmark & \error & \error \\
        zlib & Compression & \error & \checkmark & \error & ttygif & Linux & \checkmark & \error & \error \\
        lz4 & Compression & \checkmark & \error & \error & box64 & Linux & \checkmark & \error & \error \\
        libarchive & Compression & \checkmark & \error & \error & fsearch & Linux & \error & \checkmark & \error \\
        mbedtls & Crypto & \checkmark & \error & \error & uftrace & Linux & \checkmark & \error & \error \\
        libsodium & Crypto & \checkmark & \error & \error & libtree & Linux & \checkmark & \error & \error \\
        wolfssl & Crypto & \error & \checkmark & \error & toybox & Linux & \checkmark & \error & \error \\
        nettle & Crypto & \error & \checkmark & \error & tinyvm & Linux & \error & \error & \checkmark \\
        libtomcrypt & Crypto & \checkmark & \error & \error & libpcap & Linux & \error & \error & \checkmark \\
        libbcrypt & Crypto & \checkmark & \error & \error & curl & Networking & \error & \checkmark & \error \\
        tiny-AES-c & Crypto & \error & \error & \checkmark & masscan & Networking & \checkmark & \error & \error \\
        boringssl & Crypto & \checkmark & \error & \error & Mongoose & Networking & \error & \checkmark & \error \\
        tea-c & Crypto & \checkmark & \error & \error & libhv & Networking & \checkmark & \error & \error \\
        cryptopp & Crypto & \error & \checkmark & \error & wrk & Networking & \error & \error & \checkmark \\
        botan & Crypto & \error & \checkmark & \error & dsvpn & Networking & \checkmark & \error & \error \\
        openssl & Crypto & \checkmark & \error & \error & streem & Networking & \checkmark & \error & \error \\
        Tongsuo & Crypto & \checkmark & \error & \error & vlmcsd & Networking & \error & \error & \checkmark \\
        GmSSL & Crypto & \checkmark & \error & \error & acl & Networking & \checkmark & \error & \error \\
        libgcrypt & Crypto & \checkmark & \error & \error & odyssey & Networking & \checkmark & \error & \error \\
        redis & Database & \checkmark & \error & \error & massdns & Networking & \checkmark & \error & \error \\
        libbson & Database & \error & \checkmark & \error & h2o & Networking & \error & \checkmark & \error \\
        beanstalkd & Database & \checkmark & \error & \error & \parbox{2cm}{ios-webkit-debug-proxy} & Networking & \checkmark & \error & \error \\
        wiredtiger & Database & \error & \checkmark & \error & whisper.cpp & NN\tnote{2} & \checkmark & \error & \error \\
        sqlite & Database & \checkmark & \error & \error & llama2.c & NN & \checkmark & \error & \error \\
        ultrajson & DataProcessing & \error & \error & \checkmark & pocketsphinx & NN & \checkmark & \error & \error \\
        webdis & DataProcessing & \checkmark & \error & \error & lvgl & Programming & \error & \error & \checkmark \\
        jansson & DataProcessing & \checkmark & \error & \error & libui & Programming & \checkmark & \error & \error \\
        json-c & DataProcessing & \checkmark & \error & \error & quickjs & Programming & \error & \checkmark & \error \\
        libexpat & DataProcessing & \checkmark & \error & \error & flex & Programming & \checkmark & \error & \error \\
        libelf & DataProcessing & \error & \error & \checkmark & libmodbus & Security & \checkmark & \error & \error \\
        libusb & Embedded & \error & \checkmark & \error & msquic & Security & \checkmark & \error & \error \\
        wasm3 & Embedded & \checkmark & \error & \error & dount & Security & \checkmark & \error & \error \\
        rtl\_433 & Embedded & \checkmark & \error & \error & redsocks & Security & \error & \checkmark & \error \\
        can-utils & Embedded & \checkmark & \error & \error & pwnat & Security & \error & \error & \checkmark \\
        cc65 & Embedded & \error & \checkmark & \error & suricata & Security & \error & \checkmark & \error \\
        libffi & Embedded & \checkmark & \error & \error & tini & Security & \checkmark & \error & \error \\
        uhubctl & Embedded & \checkmark & \error & \error & tmux & Terminal & \checkmark & \error & \error \\
        open62541 & Embedded & \error & \checkmark & \error & sc-im & Terminal & \checkmark & \error & \error \\
        snapraid & Embedded & \checkmark & \error & \error & pspg & Terminal & \checkmark & \error & \error \\
        cglm & HPC\tnote{1} & \checkmark & \error & \error & smenu & Terminal & \checkmark & \error & \error \\
        blis & HPC & \checkmark & \error & \error & no-more-secrets & Terminal & \checkmark & \error & \error \\
        zlog & HPC & \checkmark & \error & \error & linenoise & Terminal & \error & \error & \checkmark \\
        ompi & HPC & \error & \checkmark & \error & shc & Terminal & \checkmark & \error & \error \\
        coz & HPC & \checkmark & \error & \error & hstr & Terminal & \checkmark & \error & \error \\
        ImageMagick & Image & \error & \checkmark & \error & goaccess & Terminal & \checkmark & \error & \error \\
        \bottomrule
    \end{tabular}
    \begin{tablenotes}
        \footnotesize
        \item[1] HPC stands for High Performance Computing.
        \item[2] NN stands for Neural Network.
    \end{tablenotes}
    \end{threeparttable}
    }
    \label{tab:table4}
\end{table*}

\clearpage

\end{document}